\documentclass{iau} 
\usepackage{graphicx}
\usepackage{natbib}
\usepackage{hyperref}

\providecommand{\aap}{A\&A}  
\providecommand{\aj}{AJ}      
\providecommand{\apj}{ApJ}      
\providecommand{\apjl}{ApJL}   
\providecommand{\apjs}{ApJS}   
\providecommand{\mnras}{MNRAS} 

\providecommand{\bain}{Bull.~Astron.~Inst.~Netherlands} 

\newcommand{\arcmin}{^\prime}
\newcommand{\kms}{\rm km\ s^{-1}}

\title[Dynamical vs Supernova Runaways in the SMC] 
{Dynamical vs Supernova Acceleration of OB Stars in the Small Magellanic Cloud}

\author[M. S. Oey \textit{et al.}]   
{M. S. Oey$^{1}$,
 J. Dorigo Jones$^{2}$,
 G. D. Phillips$^{1}$,
 N. Castro$^{3}$,
 M. M. Dallas$^{1,4}$,
 \and  M. Moe$^{5}$}

\affiliation{$^1$University of Michigan, Astronomy Department, Ann Arbor, MI, 48109-1107, USA
\\[\affilskip]
$^{2}$University of Colorado, Department of Astrophysical and Planetary Sciences, 2000 Colorado Ave., Boulder, CO 80309, USA
\\[\affilskip]
$^3$Leibniz-Institut f\"ur Astrophysik, An der Sternwarte, 16 D-14482, Potsdam, Germany
\\[\affilskip]
$^4$Current address:  Space Telescope Science Institute, 3700 San Martin Drive, Baltimore, MD 21218, USA
\\[\affilskip]
$^5$University of Arizona, Astronomy Department, Tucson, AZ, 85721, USA
\\[\affilskip]
}

\pubyear{2022}
\volume{361}  
\jname{Massive Stars Near and Far}
\editors{N.~St-Louis, J.~S.~Vink \& J.~Mackey, eds.}
\begin{document}

\maketitle

\begin{abstract}
We use the RIOTS4 sample of SMC field OB stars to determine the origin of massive runaways in this low-metallicity galaxy using Gaia proper motions, together with stellar masses obtained from RIOTS4 data. These data allow us to estimate the relative contributions of stars accelerated by the dynamical ejection vs binary supernova mechanisms, since dynamical ejection favors faster, more massive runaways, while SN ejection favors the opposite trend. In addition, we use the frequencies of classical OBe stars, high-mass X-ray binaries, and non-compact binaries to discriminate between the mechanisms. Our results show that the dynamical mechanism dominates by a factor of 2 -- 3. This also implies a significant contribution from two-step acceleration that occurs when dynamically ejected binaries are followed by SN kicks. We update our published quantitative results from Gaia DR2 proper motions with new data from DR3.
\end{abstract}

\firstsection 
\section{Introduction}

We know that at least 70\% of most massive stars become interacting binaries \citep[e.g.,][]{Sana2012, Moe2017}, and that close binaries and multiples generate runaway stars.  Therefore, field massive stars offer an important probe of the massive binary population.  In particular, the field OB stars in the Small Magellanic Cloud (SMC) are beautifully accessible, since they are bright and easily identified in this Milky Way satellite galaxy.  Since the SMC is a dwarf galaxy, the complete OB population can be observed.  Moreover, the SMC is metal-poor, allowing us to quantitatively characterize conditions and processes at low metallicity.

Our group carried out the Runaways and Isolated O-Type Star Spectroscopic Survey of the SMC \citep[RIOTS4;][]{Lamb2016}, which identified and spectroscopically confirmed a uniform sample of field OB stars in the SMC.  The target stars are those at least 28 pc from other OB stars in this galaxy, photometrically identified as having $B\leq 15.21$ and reddening-free $Q$-parameter $\leq -0.84$ \citep{Oey2004}.  \citet[][and these proceedings]{Vargas2020} find evidence that $\sim5$\% of the sample may have formed in situ \citep[see also][]{Oey2013, deWit2004, Pflamm2010}.  Therefore, $\sim95$\% of these field OB stars are ejected from clusters.

There are two ejection mechanisms.  Following the nomenclature of \citet{Hoogerwerf2000}, they are the dynamical ejection scenario \citep[DES; e.g.,][]{Poveda1967} and the binary supernova scenario \citep[BSS; e.g.,][]{Blaauw1961}.  The DES mechanism relies heavily on the interaction of binaries with other stars and other binaries or multiples to accelerate stars above the escape velocities of their parent clusters.  This includes also ejecting some tight binaries.  Meanwhile, the BSS mechanism requires only a single close binary system, in which the SN of the primary unbinds the remaining star, which may also receive a kick from the explosion.  On average, the DES produces faster, more massive runaways since it leverages the gravitational energy of multiple stars, while the BSS generates lower-velocity ejections of lower-mass stars.  In binaries ejected by the DES, the explosion of the primary will re-accelerate the companion, resulting in a two-step ejection \citep{Pflamm2010}.

\section{OBe stars as BSS products}

Classical OBe stars are known to be fast rotators that have generated decretion disks responsible for their characteristic Balmer emission \citep[e.g.,][]{Rivinius2013, DorigoJones2020}.  These stars most likely acquired their high rotation velocities during binary mass exchange, which also transfers angular momentum \citep[e.g.,][]{Kriz1975, Pols1991}.  Therefore, many, if not most, of these stars are likely to be post-SN systems and thus BSS products.
Following \citet{Smith2015}, we compare the spatial distributions of various populations using the distance of each star from the nearest O-star.  

\begin{figure}[ht]
\begin{center}
\includegraphics[width=5in]{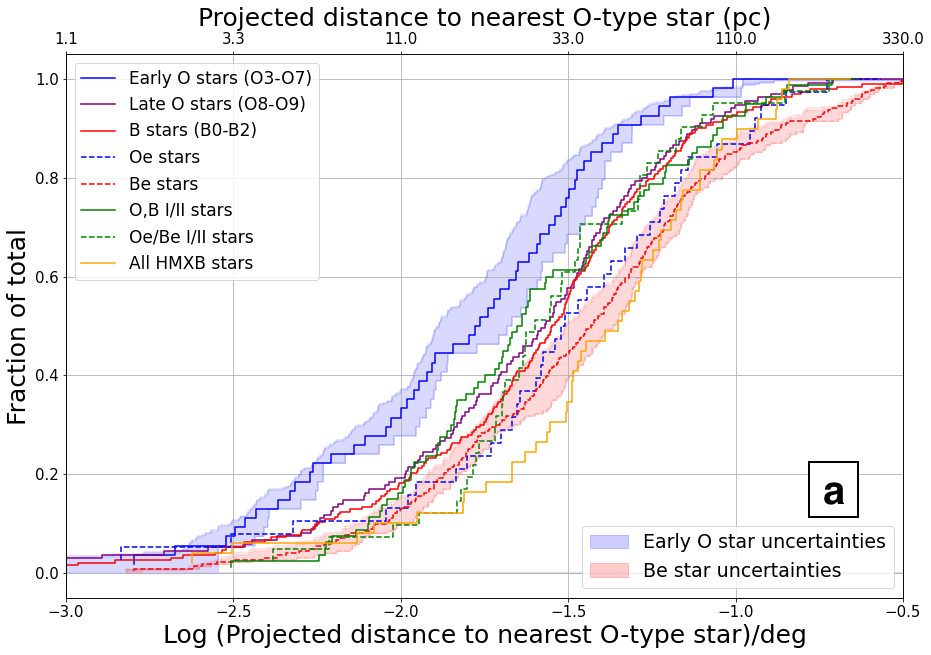} 
\caption{Cumulative distribution function of projected distances from nearest O-stars for different populations as shown.
From Dallas, Oey \& Castro (2022), in preparation. (A color version of this figure is available in the electronic version.)}
\label{fig:OBeCDFs}
\end{center}
\end{figure}

Figure~\ref{fig:OBeCDFs} shows the cumulative distribution functions of nearest O-star distances for early and late O-stars, B-stars, Oe stars, Be stars, supergiants (luminosity class I and II), and high-mass X-ray binaries (HMXBs).
As expected, early O-stars are closest to other O-stars, while late O-stars and B-stars have larger median distances since these longer-lived populations can disperse farther into the field.  The Oe and Be stars have even farther median distances, supporting the scenario that they are dominated by BSS products ejected from clusters.  In addition, the OBe distributions have a similar locus to the HMXBs, which are bona fide BSS products.  We also see that the Oe and Be distributions are indistinguishable, within the uncertainties.  This is again consistent with their mass-transfer origin, implying that their current masses are independent of their initial masses.  Thus, several lines of evidence in our data point to OBe stars largely corresponding to BSS products.  These results are discussed in more detail in Dallas \& Oey (2022, in preparation).

Thus, to first order, we can distinguish the DES and BSS products by using the field OBe stars to represent BSS products and the remaining stars to represent DES products.  We caution that these are unlikely to be exact representations of these populations, since some BSS systems may produce non-OBe stars, and the OBe phenomenon may also result from other, unrelated processes.  We also noted above that some OB stars may have formed as in-situ field stars, which would not be ejected from clusters.  However, we can use these observations to make a rough first estimate of the relative contributions of the two acceleration mechanisms.

\section{Kinematics of DES versus BSS ejections} 

\begin{figure}[ht]
\begin{center}
\includegraphics[width=4.5in]{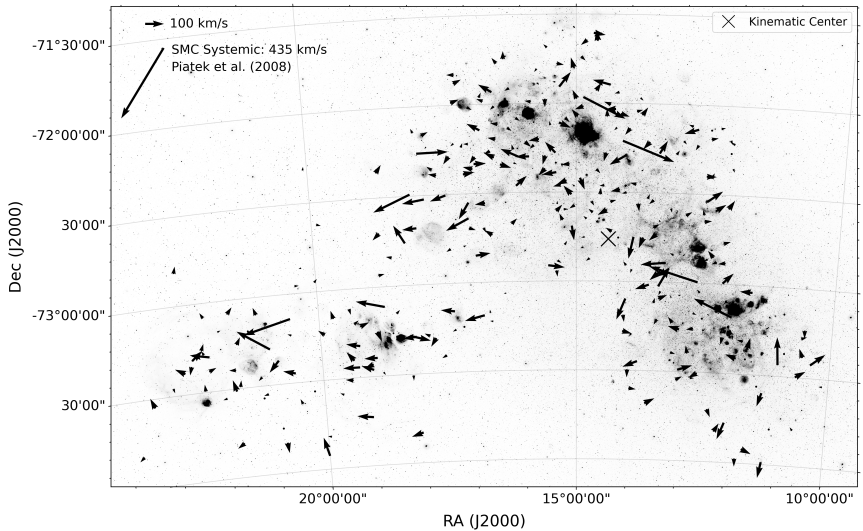} 
\caption{Map of the SMC with vectors showing residual transverse velocities \citet{Oey2018} relative to the local velocity fields within $5\arcmin$ of each RIOTS4 target star, based on Gaia DR2 data for 304 RIOTS4 stars.}
\label{fig:vectors}
\end{center}
\end{figure}

Adopting OBe stars as tracers of the BSS population, we use Gaia proper motions of the RIOTS4 sample to evaluate the relative contributions from the DES and BSS mechanisms, as well as the two-step mechanism.  Figure~\ref{fig:vectors} shows the proper motion velocity vectors for the 304 RIOTS4 stars with available DR2 Gaia data \citep{DorigoJones2020}.  These are calculated in the frame of the local velocity fields based on the blue stars having $(B-V) < 0.14$ and $M_{\rm bol}< -7.0$ from \citep{Massey2002} that are within $5\arcmin$ (90 pc) of each RIOTS4 target \citep{Oey2018}.

\begin{figure}[ht]
\vspace*{-2.0 cm}
\begin{center}
\hspace*{-0.5 cm}
\includegraphics[width=6in]{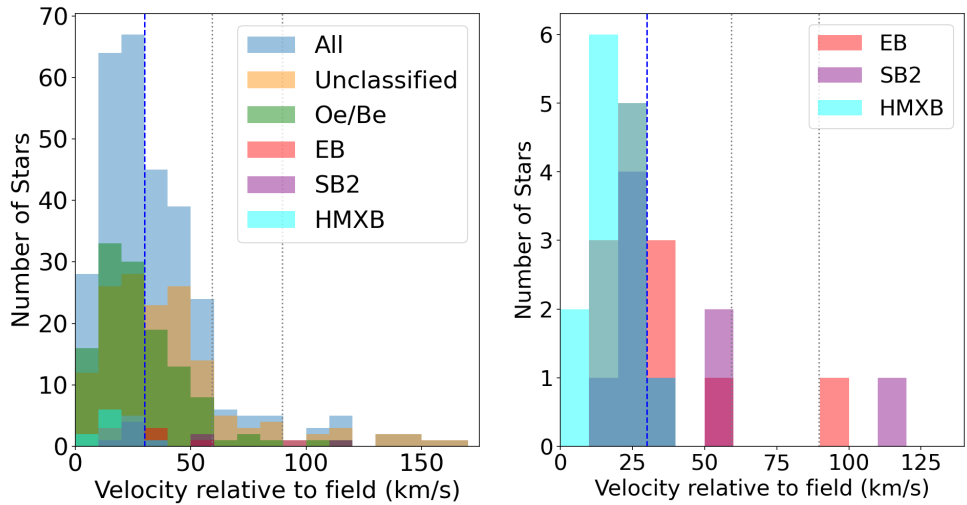} 
\vspace*{-2.0 cm}
\caption{Velocity distribution for RIOTS4 field stars and various subsets, as shown.  The right panel is a zoom showing the eclipsing binaries (EB), double-lined spectroscopic binaries (SB2), and high-mass X-ray binaries (HMXB).  ``Unclassified" objects are those that do not belong to any of the other subsets.  The dashed vertical line shows a nominal 30 $\kms$ value, and the dotted lines show 1 and 2 standard deviations above the median of $28.6\ \kms$.
(A color version of this figure is available in the electronic version.)}
\label{fig:veldistrib}
\end{center}
\end{figure}

We show updated velocity distributions from Gaia EDR3 for 299 RIOTS4 stars in Figure~\ref{fig:veldistrib}.  We see that the OBe stars, which are putative BSS products, are indeed at lower median velocities than the remaining stars, which conversely must be a population dominated by DES ejections.  Eclipsing binaries (EBs) and double-lined spectroscopic binaries (SB2s) are non-compact, pre-SN binaries that  are thus bona fide tracers of the DES mechanism.  These can be compared to the HMXBs, which as mentioned above, are similarly tracers of BSS ejections.  We likewise see that the HMXBs are restricted to lower velocities while the EBs and SB2s have distributions extending to much higher velocities, mirroring what we see for the OBe versus non-OBe stars.

\begin{figure}[ht]
\begin{center}
\includegraphics[width=2.5in]{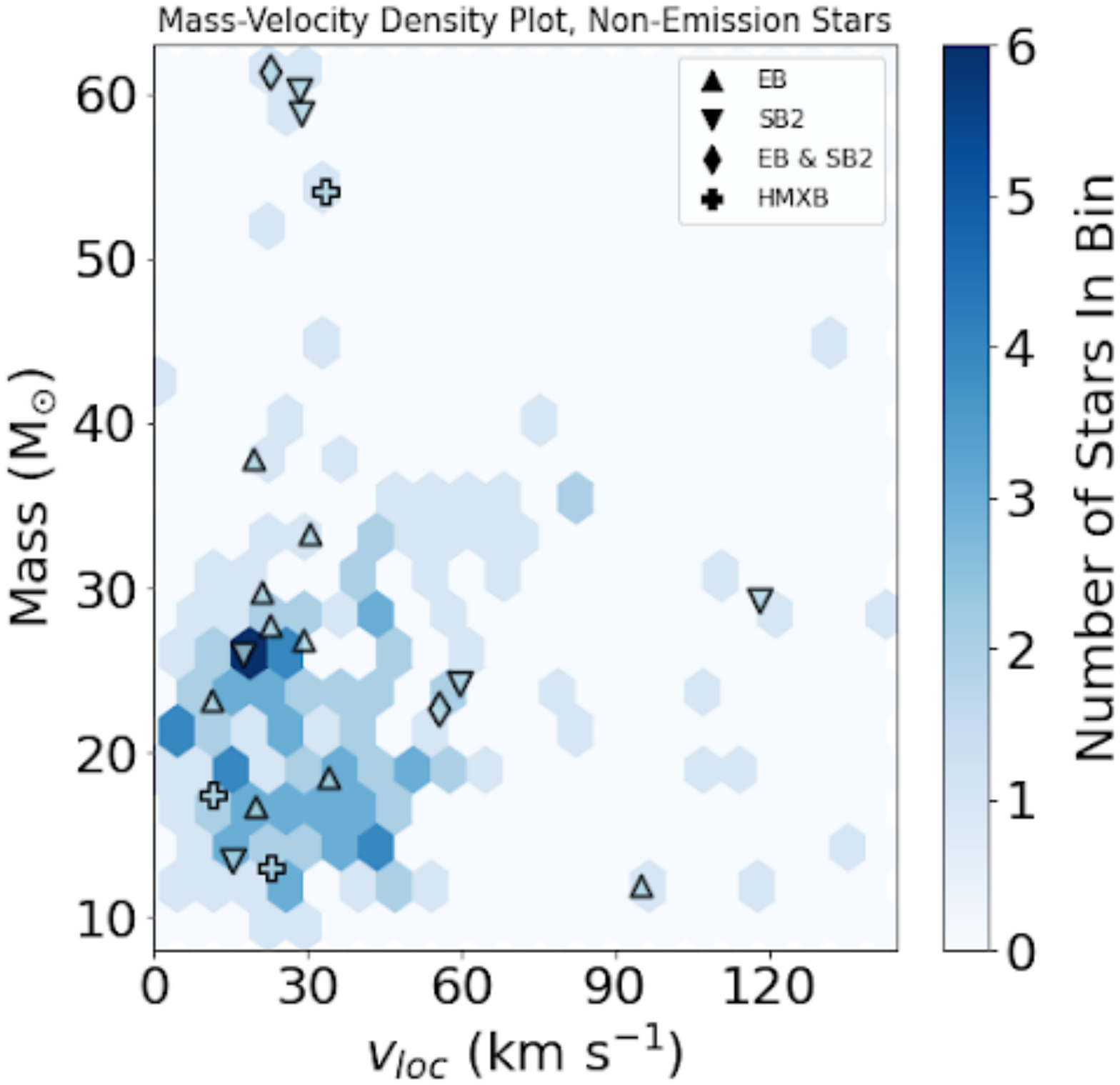}~~
\includegraphics[width=2.5in]{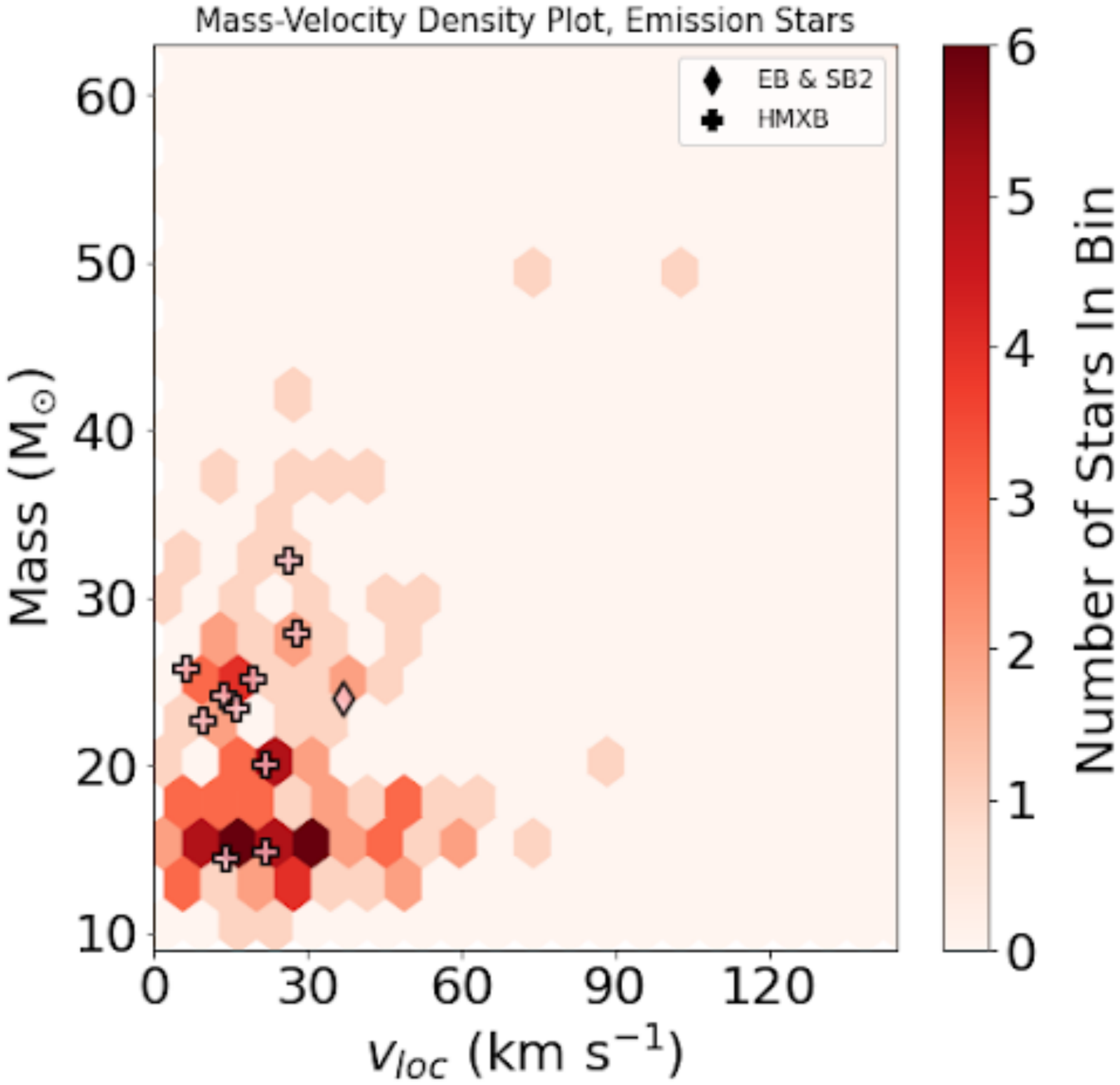} 
\vspace*{-1.0 cm}
\caption{Distributions of mass vs transverse velocity for 283 stars in our RIOTS4 field star sample with mass determinations.  The left panel shows non-OBe stars and the right panel shows OBe stars. Individual EBs, SB2s, and HMXBs are overplotted as shown.}
\label{fig:density}
\end{center}
\end{figure}

Figure~\ref{fig:veldistrib} shows that the OBe velocity distribution extends to relatively high velocities, well in excess of 30 $\kms$, whereas BSS products are largely expected to remain at ``walkaway" velocities below this value.  This may result from Gaia errors, which are asymmetric and tend to be positive, rather than negative.  We find that the median velocity of the entire distribution decreases about 25\%, from 39 $\kms$ to 29 $\kms$, between DR2 and EDR3.  However, another, real effect is that some OBe stars have been doubly accelerated, originating as non-compact DES ejections that have subsequently experienced a SN explosion.  These are the ``two-step" ejections \citep{Pflamm2010}.

Figure~\ref{fig:density} shows density plots of the non-OBe and OBe populations for stellar mass vs velocity.  Their tracers, the EB and SB2 populations (DES), and the HMXBs (BSS), are overplotted.  
The data are again consistent with the expectation that DES products on average accelerate more massive runaways to faster velocities, while BSS products show opposite trends.  The tracer populations are also consistent with these patterns.

Based on the DR2 proper motion data, \citet{DorigoJones2020} find that on the order of 26\% of the SMC OB population corresponds to DES products, including $\sim8$\% non-compact binaries; while $\sim 9$\% are BSS products and perhaps an additional $\sim 2$\% are two-step ejections.  These values are model-dependent and rely on parameters from \break N-body simulations for the DES mechanism by \citet{Oh2016} and binary population synthesis models by \citet{Renzo2019} for the BSS. The total frequencies are larger than the SMC field population because they include ejections that do not meet the criteria for RIOTS4 field stars \citep[see Section 5.1 of][for details]{DorigoJones2020}.
We caution that the measured velocities changed substantially between Gaia DR2 and EDR3, as noted above. Thus, our derived frequencies have substantial uncertainties and represent only a first crude estimate of the relative contributions of the different ejection mechanisms.  We are currently in the process of updating these calculations (Phillips et al., in preparation).  However, the ability to make such estimates based on empirical data promises major advances in our understanding of massive binary populations and cluster dynamics.

\section*{Acknowledgements}

This work was supported in part by NSF grant AST-1514838 to M.S.O.  N. C. acknowledges funding from the Deutsche Forschungsgemeinschaft
(DFG), CA 2551/1-1.

\bibliographystyle{apj}

\begin{discussion}

\discuss{Shenar}{
Some later Be stars could originate from mass donors that are less massive and would not have exploded.  Have you considered this in your velocity distributions?
 }

\discuss{Oey}{
No, we have not included this effect.  Given that binaries tend to have high mass ratios, the contribution from such objects should be relatively small. There are also others that we ignored, for example, not all stars that undergo mass transfer will be seen as Be stars and similarly, there may also be a significant number of Be stars that originate from a completely unrelated mechanism.  So we stress that our results are simply a first rough estimate of the relative contributions of the acceleration mechanisms.
 }

\end{discussion}

\end{document}